%% file: BESIII_draft_D0p_b1ev_submit.tex
\documentclass[aps,prl,twocolumn,showpacs,amsmath,amssymb]{revtex4-1}
\usepackage{amsmath}
\usepackage{graphicx}
\usepackage{subfigure}
\usepackage{epstopdf}
\usepackage{color}
\usepackage{multirow}
\usepackage{setspace}
\usepackage{overpic}
\usepackage{amssymb}
\usepackage[ bookmarksnumbered, pdfstartview=FitH,colorlinks,urlcolor=blue, citecolor=blue,linkcolor=blue] {hyperref}
\usepackage{lineno}
\usepackage{bm}
\usepackage{rotating}
\usepackage[utf8]{inputenc}
\let\oldequation\equation
\let\oldendequation\endequation

\renewenvironment{equation}
  {\linenomathNonumbers\oldequation}
  {\oldendequation\endlinenomath}

 \newcommand{\BESIIIorcid}[1]{\href{https://orcid.org/#1}{\hspace*{0.1em}\raisebox{-0.45ex}{\includegraphics[width=1em]{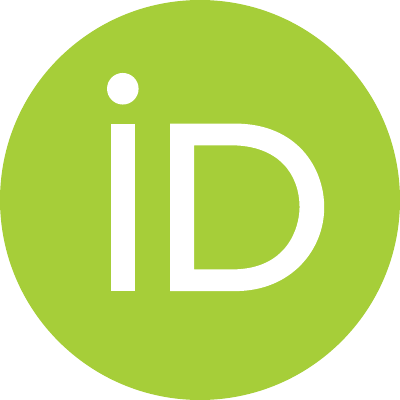}}}}
\begin{document}

\title{\bf \boldmath
Observation of the singly Cabibbo suppressed decay $D^0\to b_1(1235)^- e^+\nu_e$ and evidence for $D^+\to b_1(1235)^0 e^+\nu_e$}

\input{authorlist_2024-04-30}

\begin{abstract}
By analyzing a data sample of $e^+e^-$ collisions with center-of-mass energy $\sqrt{s}=3.773$~GeV, corresponding to an integrated luminosity of $7.9~\rm {fb}^{-1}$ collected with the BESIII detector operating at the BEPCII collider,
we study semileptonic decays of the $D^{0(+)}$ mesons into the axial-vector meson $b_1(1235)$ via the decay $b_1(1235)\to \omega\pi$.
The decay $D^0\to b_1(1235)^-e^{+}\nu_{e}$ is observed with a significance of 5.2$\sigma$ after considering systematic uncertainty, while evidence for the decay $D^+\to b_1(1235)^0 e^+\nu_e$ is obtained with a 3.1$\sigma$  significance.  The product branching fractions  are determined to be ${\mathcal B}(D^0\to b_{1}(1235)^-e^{+}\nu_{e})\times {\mathcal B} (b_1(1235)^-\to \omega \pi^-) = (0.72\pm0.18^{+0.06}_{-0.08})\times10^{-4}$  and ${\mathcal B}(D^+\to b_{1}(1235)^0e^{+}\nu_{e})\times {\mathcal B} (b_1(1235)^0~\to
\omega \pi^0) = (1.16\pm0.44\pm0.16)\times10^{-4}$, where the first uncertainties are  statistical and the second systematic. The ratio of their partial decay widths is determined to be   $\frac{\Gamma(D^0\to b_{1}(1235)^-e^{+}\nu_{e})}{2\Gamma(D^+\to b_{1}(1235)^0e^{+}\nu_{e})}=0.78\pm0.19^{+0.04}_{-0.05}$, which is consistent with unity, predicted by isospin invariance, within  uncertainties.

\end{abstract}

\maketitle

\oddsidemargin  -0.2cm
\evensidemargin -0.2cm

Experimental investigations of light hadron spectroscopy in semileptonic decays of the $D$ mesons ($D$ denotes $D^0$ or $D^+$) 
can be used to shed light on the role of non-perturbative strong interactions in weak decays, thereby aiding in uncovering the internal
structure of the hadrons involved.
The first quantitative predictions for the partial widths of
various semileptonic $D$ decays into $S$- or $P$-wave light meson states came from
the quark model developed by Isgur, Scora, Grinstein, and Wise (namely ISGW)~\cite{isgw}. 
This model was later updated to include constraints from heavy quark
symmetry, hyperfine distortions of wave functions, and form factors
with more realistic behavior at high recoil masses~(ISGW2)~\cite{isgw2}.
To date, studies of Cabibbo-suppressed semileptonic $D$ decays
are not as advanced as their Cabibbo-favored counterparts, which have been extensively studied both theoretically and experimentally~\cite{pdg2024,Ke:2023qzc,Li:2021iwf}. 
In general, these decays, which are mediated by the quark level process $c\to de^+\nu_e$,
are expected to be dominated by the ground state pseudoscalar and vector  mesons.
Due to limited phase space, 
heavier mesons, such as $P$-wave states or the first radial excitations
of the $d\bar{u}$ and $d\bar{d}$ mesons, 
are less likely to be produced. 
Among the heavier mesons, the most promising to be produced is the $P$-wave $b_1$ meson, which is accessible via $D\to b_1 e^+\nu_e$. 
Throughout this Letter, $b_1$ denotes $b_1(1235)$.
The ISGW2 model predicts the branching fractions of $D^0\to b_1^-e^+\nu_e$ and $D^+\to b_1^0e^+\nu_e$
to be $1.08\times 10^{-4}$ and $1.32\times 10^{-4}$~\cite{isgw2}, respectively.
The covariant light-front quark model predicts the branching fraction of
$D^+\to b_1^0e^+\nu_e$ to be $(7.4\pm0.7)\times 10^{-5}$~\cite{cheng}.
The first measurements of the branching fractions of the $D^{0(+)}\to b_{1}^{-(0)} e^+ \nu_e$ decays
help to validate these quark model predictions.

The axial-vector meson $b_1$, which is identified as a $^1P_1$ state in the quark model, was first observed by the HBC Collaboration in 1963~\cite{MA:1963}.
Before 1994, it was  usually studied in $p\bar p$~\cite{ppbar:1967,ppbar:1969}, $\gamma p$~\cite{gammap}, and $p\pi$~\cite{ppi:1974_1,ppi:1974_2,ppi:1977,ppi:1981,ppi:1984,ppi:1984_1,ppi:1991,ppi:1992,ppi:1993} reactions, focusing on mass and width measurements. 
Later, it was widely observed in hadronic decays of $D$, $B$, and charmonium states~\cite{pdg2024}.
Besides the $q\bar q$ configuration, there are many different theoretical models for the
nature of the $b_1$, such as a mixture of an ordinary $q\bar q$ meson and a hybrid $q\bar q g$ meson~\cite{Ishida:1991nz}, a molecular state which is dynamically generated by the pseudoscalar-vector
meson interaction~\cite{Lutz:2003fm,Roca:2005nm,Zhou:2014ila,Liang:2019vhf,Clymton:2023txd}, especially with strong couplings to the $K{\bar K}^*$ channel, etc. 
Historically, theorists and experimentalists usually considered the $\omega\pi$ decay mode to be dominant for the axial-vector $b_1$ meson. However, the next-to-leading order (NLO) chiral perturbation calculations~\cite{Zhou:2014ila} predict that the $\phi\pi$ channel becomes the dominant contribution under specific molecular-state assumptions. 
The $b_1$ mesons produced in semileptonic $D$ decays offer an ideal opportunity to validate these predictions, thereby gaining insight into their nature~\cite{bes3-white-paper}.
Furthermore, compared to high backgrounds and complex interference patterns found in hadronic processes~\cite{pdg2024},
the semileptonic $D$ meson decays are theoretically cleaner  because leptons are not involved in strong interaction~\cite{M.A}. The verification 
of theoretical calculations of semileptonic $D$ decays into the $b_1$  help to constrain the theoretical calculations in the decays of the $\tau$~\cite{tau2:2018,tau1:2020}, $B$~\cite{B:2012,B:2012ew,B:2017}, $D$~\cite{D:2019,D:2022}, and charmonium states with $b_1$ involved in the final states~\cite{jpsi:2001,jpsi:2004,jpsi:2019}.

To date, there is limited experimental information available on $D \to b_1e^+\nu_e$ decays. Only BESIII has previously searched for $D\to b_{1}\ell^+\nu_\ell$,
using a portion of the same data used in this analysis,
but no significant signal was obtained~\cite{b1ev}. This Letter reports the first observation of $D^0\to b_1^- e^+\nu_e$ and the first evidence for  $D^+\to b_1^0 e^+\nu_e$ with $b_1^{-(0)}\to\omega\pi^{-(0)}$,
by analyzing 7.9~fb$^{-1}$ of $e^+e^-$ collision data taken at $\sqrt s=3.773$~GeV~\cite{lumi}.
Throughout this Letter, charge conjugate channels are always implied.

Details about the design and performance of the BESIII detector are given in
Ref.~\cite{BESIII}. The end-cap time-of-flight (TOF)  system was upgraded in 2015 using multigap
resistive plate chamber technology, providing a time
resolution of 60 ps~\cite{etof}. Approximately 63\% of the data
used here was collected after this upgrade. Simulated samples produced with the {\sc geant4}-based~\cite{geant4} Monte Carlo (MC) package which
includes the geometric description of the BESIII detector and the
detector response, are used to determine the detection efficiency
and to estimate the backgrounds. The simulation includes the beam
energy spread and initial state radiation (ISR) in the $e^+e^-$
annihilations modeled with the generator {\sc kkmc}~\cite{kkmc}.
The inclusive MC samples consist of the production of $D\bar{D}$
pairs with consideration of quantum coherence for all neutral $D$
modes, the non-$D\bar{D}$ decays of the $\psi(3770)$, the ISR
production of the $J/\psi$ and $\psi(3686)$ states, and the
continuum processes.
The known decay modes are modeled with {\sc
evtgen}~\cite{evtgen} using branching fractions taken from the
Particle Data Group
~\cite{pdg2024}, and the remaining unknown charmonium decays
are estimated with {\sc
lundcharm}~\cite{lundcharm}. Final state radiation
from charged final state particles is incorporated using the {\sc
photos} package~\cite{photos}.
The $D \to b_1 e^+\nu_e$ signal MC events are simulated using the ISGW2 model~\cite{isgw2},
where the $b_1$ is parameterized by a relativistic Breit-Wigner function
with mass and width fixed to the world-average values of $1229.5\pm 3.2$ MeV/$c^{2}$ and $142\pm9$ MeV, respectively~\cite{pdg2024}.

At $\sqrt s=3.773$ GeV, the $D$ and $\bar D$ mesons are produced in pairs without accompanying particles in the final state.
Because of this, one can study semileptonic $D$ decays with the double-tag~(DT) method.
Single-tag~(ST) $\bar D^0$ mesons are first reconstructed using the hadronic decay modes
$\bar D^0\to K^+\pi^-$, $K^+\pi^-\pi^0$, $K^+\pi^-\pi^-\pi^+$, $K_{S}^0\pi^+\pi^-$, $K^+\pi^-\pi^0\pi^0$, and $K^+\pi^+\pi^-\pi^-\pi^0$;
while ST $D^-$ mesons are reconstructed via the decay modes
$D^-\to K^{+}\pi^{-}\pi^{-}$,
$K^0_{S}\pi^{-}$, $K^{+}\pi^{-}\pi^{-}\pi^{0}$, $K^0_{S}\pi^{-}\pi^{0}$, $K^0_{S}\pi^{+}\pi^{-}\pi^{-}$,
and $K^{+}K^{-}\pi^{-}$.
Semileptonic $D$
candidates are then reconstructed from the residual  tracks and showers not used in the tag selection.
Candidate events in which the $D$ decays into $b_1e^+\nu_e$ and the $\bar D$ decays into a tag mode are
called DT events.

Because the branching fraction of the $b_1 \to \omega \pi$ decay is unknown,
the product branching fractions of the decay $D\to b_1e^+\nu_e$ (${\mathcal B}_{\rm SL}$) and its subsequent decay $b_1\to\omega\pi$ (${\mathcal B}_{b_1}$) are determined from
\begin{equation}
\label{eq:bf}
{\mathcal B}_{\rm SL}\cdot {\mathcal B}_{b_1} = \frac{N_{\rm DT}}{N^{\rm tot}_{\rm ST}\cdot\bar \varepsilon_{\rm SL}\cdot {\mathcal B}_{\rm sub}},
\end{equation}
where $N_{\rm ST}^{\rm tot}=\Sigma_{i}N^{i}_{\rm ST}$ and $N_{\rm DT}$ are the total ST and
DT yields after summing over all tag modes $i$;
${\mathcal B}_{\rm sub}={\mathcal B}_{\omega}\cdot {\mathcal B}_{\pi^0}$ for $D^0\to b_1^-e^+\nu_e$ and
${\mathcal B}_{\rm sub}={\mathcal B}_{\omega}\cdot {\mathcal B}^2_{\pi^0}$  for $D^+\to b_1^0e^+\nu_e$,
in which ${\mathcal B}_{\omega}$ and ${\mathcal B}_{\pi^0}$ are the branching fractions of  $\omega\to\pi^+\pi^-\pi^0$ and $\pi^0\to \gamma\gamma$, respectively;
and $\bar\varepsilon_{\rm SL}=\Sigma_{i}\frac{N^{i}_{\rm ST}}{N^{\rm tot}_{\rm ST}}\frac{\epsilon^i_{\rm DT}}{\epsilon^i_{\rm ST}}$ is the average signal efficiency of reconstructing $D\to  b_1e^+\nu_e$ in the presence of an ST $D$, where $\epsilon^i_{\rm ST}$ and $\epsilon^i_{\rm DT}$ are the ST and DT efficiencies for the $i$-th tag mode, respectively.

All charged tracks detected in the multilayer drift chamber (MDC)  must be within a polar angle ($\theta$) range of $|\rm{cos\theta}|<0.93$, where $\theta$ is defined with respect to the $z$-axis,
which is the symmetry axis of the MDC.
Except for charged tracks from $K^0_S$ decays,
they must originate from an interaction region defined by $|V_{xy}| <$ 1\,cm and
$|V_{z}| <$10\,cm. Here, $|V_{xy}|$ and $|V_{z}|$ denote the distances
of closest approach of the reconstructed track to the
interaction point (IP) in the $xy$ plane and the $z$ direction
(along the beam), respectively.
Particle identification~(PID) for charged tracks combines measurements of the specific ionization energy loss in the MDC~(d$E$/d$x$) and the flight time in the TOF to form confidence levels for pion and kaon hypotheses~(${\rm CL}_\pi$  and ${\rm CL}_K$).
Charged tracks with  ${\rm CL}_K>{\rm CL}_\pi$ and ${\rm CL}_\pi>{\rm CL}_K$ are assigned as kaons and pions, respectively.

Each $K_{S}^0$ candidate is reconstructed from two oppositely charged tracks satisfying $|V_{z}|<20$~cm.
The two charged tracks are assigned as $\pi^+\pi^-$ without imposing further PID criteria.
They are constrained to originate from a common vertex, and required to have  an invariant mass within $(0.487,0.511)$~GeV/$c^2$.
The decay length of the $K^0_S$ candidate is required to be greater than
twice the vertex resolution away from the IP.
The quality of the vertex fits (primary-vertex fit and secondary-vertex fit) is ensured by a
requirement of $\chi^2 < 100$.

Photon candidates are selected using showers in the electromagnetic calorimeter (EMC).
The deposited energy of each shower must be more than 25~MeV in the barrel region ($|\!\cos \theta|< 0.80$) and more than 50~MeV in the end-cap region ($0.86 <|\!\cos \theta|< 0.92$).
To exclude showers associated with  charged tracks,
the angle subtended by the EMC shower and the position of the closest charged track at the EMC
must be greater than $10^\circ$  as measured from the IP.
To suppress electronic noise and showers unrelated to the event, the difference between the EMC time and the event start time is required to be within
[0, 700]\,ns.
For $\pi^0$ candidates, the invariant mass of the photon pair is required to be within $(0.115,\,0.150)$\,GeV$/c^{2}$. To improve the momentum resolution, a   one-constraint~(1C) kinematic fit to the known  $\pi^{0}$ mass~\cite{pdg2024} is imposed on the photon pair. The four-momentum of the $\pi^0$ candidate updated by the 1C fit is kept for  further analysis.

To separate the ST $\bar D$ mesons from combinatorial backgrounds, we define the energy difference $\Delta E\equiv E_{\bar D}-E_{\rm {beam}}$ and the beam-constrained mass $M_{\rm BC}\equiv\sqrt{E_{\rm {beam}}^{2}/c^{4}-|\vec{p}_{\bar D}|^{2}/c^{2}}$, where $E_{\rm {beam}}$ is the beam energy, and $E_{\bar D}$ and $\vec{p}_{\bar D}$ are the total energy and momentum of the $\bar D$ candidate in the $e^+e^-$ center-of-mass frame, respectively.
If there is more than one $\bar D$ candidate in a given ST mode, the one with the minimal $|\Delta E|$ is kept for further analysis.
The rates of multiple combinations for tag modes containing $\pi^0$ in the final state range between (8.7-29.0)\%, while those for other tag modes range between (0.1-7.3)\%.
The $\Delta E$ requirements and ST efficiencies for the different tag modes are listed in Table~\ref{ST:realdata}.

\begin{figure}[htbp]\centering
	\includegraphics[width=1.0\linewidth]{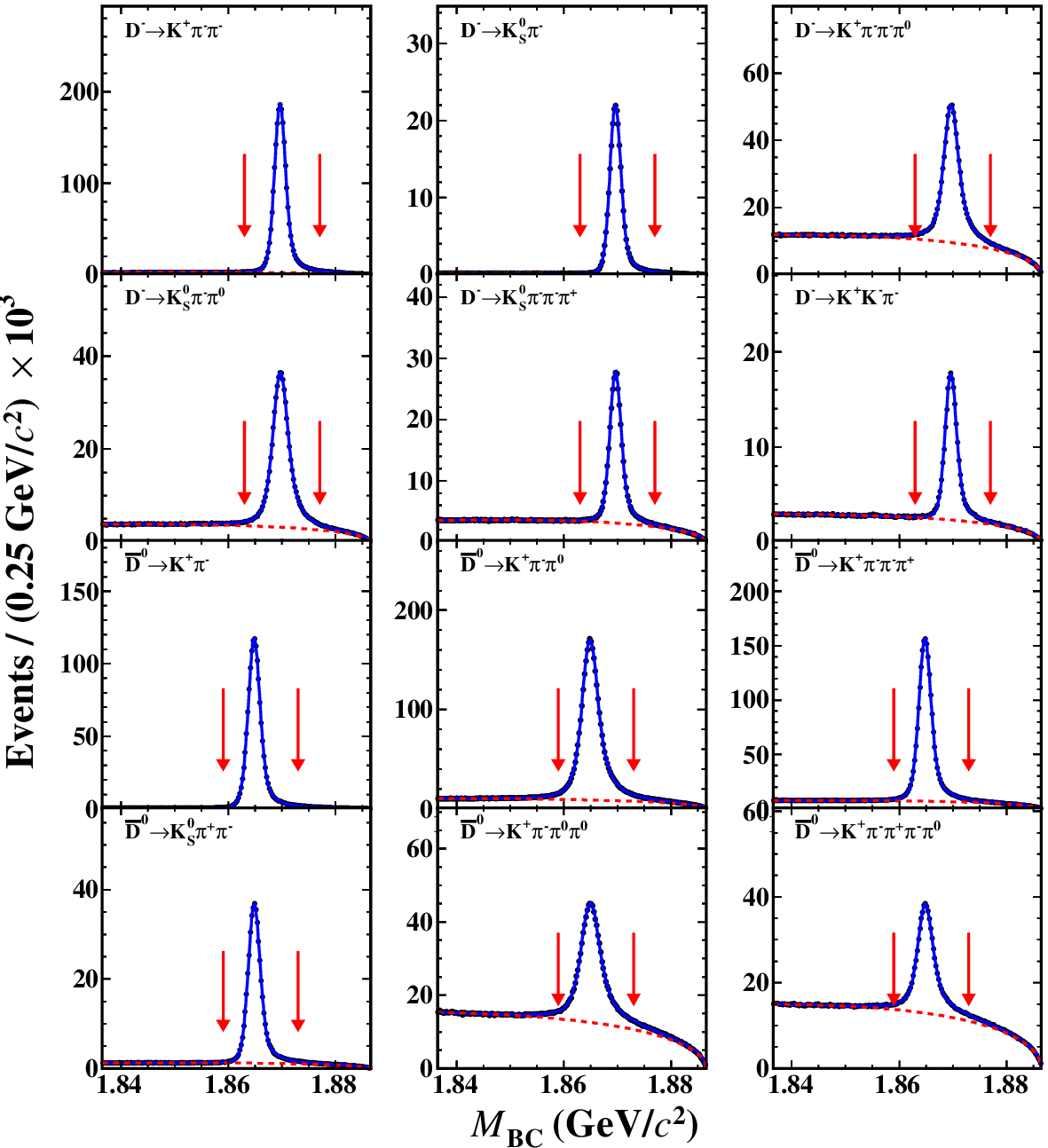}
	\caption{
		Fits to the $M_{\rm BC}$ distributions of the ST $\bar D$ candidates for different tag modes.
The dots with error bars are data, the blue curves are the best fits, and the red dashed curves are the fitted combinatorial background shapes. The pair of red arrows are the $M_{\rm BC}$ signal windows.}\label{fig:datafit_Massbc}
\end{figure}

The ST yields are extracted by performing unbinned maximum likelihood fits to individual $M_{\rm BC}$ distributions. In the fit, the signal shape is derived from the MC-simulated signal shape convolved with a double-Gaussian function to consider the resolution difference between data and MC simulation.
The background shape is described by the ARGUS function~\cite{argus}, with the endpoint parameter fixed at 1.8865~GeV/$c^{2}$ corresponding to $E_{\rm beam}$.
Figure~\ref{fig:datafit_Massbc} shows the fits to the $M_{\rm BC}$ distributions of the accepted ST candidates in data for different tag modes. The candidates with $M_{\rm BC}$ within $(1.859,1.873)$ GeV/$c^2$ for $\bar D^0$ tags and $(1.863,1.877)$ GeV/$c^2$ for $D^-$ tags are kept for further analysis. Summing over the tag modes gives the total yields of ST $\bar D^0$ and $D^-$ mesons to be $(7896.0 \pm 3.4_{\rm stat})\times 10^3$ and $(4149.9\pm2.3_{\rm stat})\times 10^3$, respectively.

\begin{table}
	\renewcommand{\arraystretch}{1.2}
	\centering
	\caption {The $\Delta E$ requirements, the measured ST $\bar D$ yields, and the ST efficiencies ($\epsilon_{\rm ST}^{i}$) for different tag modes. The uncertainties are statistical only.}
	\scalebox{0.87}{
		\begin{tabular}{cccc}
			\hline
			\hline
			Tag mode & $\Delta E$~(GeV)  &  $N^{i}_{\rm ST}~(\times 10^3)$  &  $\epsilon^i_{\rm ST}~(\%)$       \\\hline
			$\bar D^{0}\to K^{+}\pi^{-}$                           & $(-0.027,0.027)$&$1449.3\pm1.2$&     $65.34\pm0.01$ \\
			$\bar D^{0}\to K^{+}\pi^{-}\pi^{0}$                   & $(-0.062,0.049)$&$2913.1\pm2.0$&      $35.59\pm0.01$ \\
			$\bar D^{0}\to K^{+}\pi^{-}\pi^{-}\pi^{+}$            & $(-0.026,0.024)$ &$1944.1\pm1.5$&     $40.83\pm0.01$ \\
			$\bar D^{0} \to K_{S}^{0} \pi^{+} \pi^{-}$        &$(-0.024,0.024)$& $447.6\pm0.7$ &$37.49\pm0.01$\\
			$\bar D^{0} \to K^{+} \pi^{-} \pi^{0} \pi^{0}$   &$(-0.068,0.053)$& $690.6\pm1.3$ &$14.83\pm0.01$\\
			$\bar D^{0} \to K^{+} \pi^{+} \pi^{-} \pi^{-} \pi^{0}$&$(-0.057,0.051)$& $450.9\pm1.1$&$16.17\pm0.01$\\
			\hline
			$D^{-}\to K^{+} \pi^{-}\pi^{-}$                       & $(-0.025,0.024)$&$2164.0\pm1.5$&     $51.17\pm0.01$ \\
			$D^{-}\to K_S^{0} \pi^{-}$                            & $(-0.025,0.026)$&$250.4\pm0.5$&      $50.74\pm0.02$ \\
			$D^{-}\to K^{-}\pi^{-}\pi^{+}\pi^{0}$                 & $(-0.057,0.046)$&$689.0\pm1.1$&      $25.50\pm0.01$ \\
			$D^{-}\to K_S^{0} \pi^{-}\pi^{0}$                     & $(-0.062,0.049)$&$558.4\pm0.9$&      $26.28\pm0.01$ \\
			$D^{-}\to K_S^{0} \pi^{-}\pi^{-}\pi^{+}$              & $(-0.028,0.027)$& $300.5\pm0.6$&     $29.01\pm0.01$ \\
			$D^{-}\to K^{+}K^{-}\pi^{-}$                          & $(-0.024,0.023)$&$187.3\pm0.5$&      $41.06\pm0.02$ \\
			\hline
			\hline
		\end{tabular}
	}
	\label{ST:realdata}
\end{table}


In the presence of the tagged $\bar D$ candidates,
$D^0\to b_1^- e^+\nu_e$ and $D^+\to b_1^0 e^+\nu_e$ candidates are selected from the residual tracks and photons not used in the tag reconstruction.
The selection criteria of charged and neutral pions are the same as those used in the tag selection.
For the selected candidates  $D^0\to \pi^+\pi^-\pi^-\pi^0e^+\nu_e$ and $D^+\to \pi^+\pi^-\pi^0\pi^0 e^+\nu_e$,
two possible $\pi^+\pi^-\pi^0$ combinations can form the $\omega$ candidate.
In order to veto backgrounds from $D\to a_0(980)e^+\nu_e$, the invariant masses of both $\pi^+\pi^-\pi^0$ combinations ($M_{\pi^+\pi^-\pi^0}$) are
required to be greater than 0.6 GeV/$c^2$.
This rejects almost all of the $D\to a_0(980)e^+\nu_e$ background events. One  candidate is kept for further analysis if either of the $\pi^+\pi^-\pi^0$ combinations falls in  the $\omega$ mass signal region $M_{\pi^+\pi^-\pi^0}\in$  (0.757,0.807)~GeV/$c^2$. Events in the $\omega$  sideband region,  defined as $M_{\pi^+\pi^-\pi^0} \in$    (0.697,0.742) and (0.822,0.867) GeV/$c^2$, are  used to study potential  combinatorial background in the $\omega$ signal region.
To reject background events from $D^{0(+)}\to \bar K_1(1270) [\to K_S^0\pi^{+(0)}\pi^{-(0)}] e^+\nu_e$,
we require the invariant masses of all $\pi^+\pi^-$ ($\pi^0\pi^0$) combinations satisfy $|M_{\pi^+\pi^-(\pi^0\pi^0)}-M_{K_{S}^{0}}|>0.008$ GeV/$c^2$.

Identification of $e^+$ candidates is performed with combined d$E$/d$x$, TOF, and EMC information. Based on these, we calculate confidence levels for the positron, pion, and kaon hypotheses~(${\rm CL}_e$, ${\rm CL}_\pi$, and ${\rm CL}_K$).
Charged tracks satisfying  ${\rm CL}_e>0.001$ and ${\rm CL}_e/({\rm CL}_e+{\rm CL}_\pi+{\rm CL}_K)>0.8$ are assigned as $e^+$ candidates.
To further reject background events from misidentified hadrons and muons,
the deposited energy of any $e^+$ candidate in the EMC must be greater than 0.8 times its momentum reconstructed in the MDC.

The invariant mass of the $b_1e^+$ system~($M_{b_1e^+}$) is required to be less than 1.82~GeV/$c^2$ to  suppress  peaking background contributions from the decay $D\to b_1^{-(0)}\pi^+$.
To suppress backgrounds with extra photons or $\pi^0$s,   the maximum energy of any extra
photon  that has not been used in the event
selection~($E^{\rm max}_{\rm extra ~\gamma}$) must be less than 0.30~GeV, and
 there must be
no additional $\pi^0$ candidates~($N^{\pi^{0}}_{\rm extra}$) in the candidate event. In addition, the opening angle
between the missing momentum and the most energetic
unused shower when found, $\theta_{\gamma,\rm miss}$, is required to satisfy $\cos\theta_{\gamma,\rm miss}< 0.3$ and $\cos\theta_{\gamma,\rm miss}< 0.4$ for $D^0\to b_1^- e^+\nu_e$ and $D^+\to b_1^0 e^+\nu_e$, respectively.
To suppress backgrounds due
to final state radiation, the angle between the direction
of the radiative photon and the $e^+$  momentum
is required to be greater than 0.20 radians.

The presence of the missing neutrino is inferred using a kinematic variable defined as
\begin{equation}
U_{\rm miss}\equiv E_{\rm miss}-|\vec{p}_{\rm miss}|\cdot c,
\end{equation}
with
\begin{equation}
E_{\rm {miss}}\equiv E_{\rm {beam}}-E_{b_1}-E_{e^{+}},
\end{equation}
and
\begin{equation}
\vec{p}_{\rm {miss}}\equiv\vec{p}_{D}-\vec{p}_{b_1}-\vec{p}_{e^{+}},
\end{equation}
where $E_{b_1\,(e^+)}$ and $\vec{p}_{b_1\,(e^+)}$ are the measured energy and momentum of the $b_1$\,($e^+$) candidates in the $e^+e^-$ center-of-mass frame, respectively. We calculate $\vec{p}_{D}\equiv-\hat{p}_{\bar D}\cdot \sqrt{E_{\rm {beam}}^{2}/c^{2}-m_{\bar D}^{2}\cdot c^{2} }$, where
$\hat{p}_{\bar D}$ is a unit vector in the momentum direction of the ST $\bar D$ meson and $m_{\bar D}$ is the known $\bar D$ mass~\cite{pdg2024}.
The beam energy and the known $D$ mass are used to determine the magnitude of the momentum of the ST $D$ meson
in order to improve the  $U_{\rm {miss}}$ resolution.
The signal candidates are expected
to peak around zero in the $U_{\rm  miss}$ distribution and near the invariant mass of the $b_1$ in the $\omega\pi$ mass spectrum ($M_{\omega\pi}$).

\begin{figure}[htbp]
	\hspace{-0.5cm}
\includegraphics[width=1.0\linewidth]{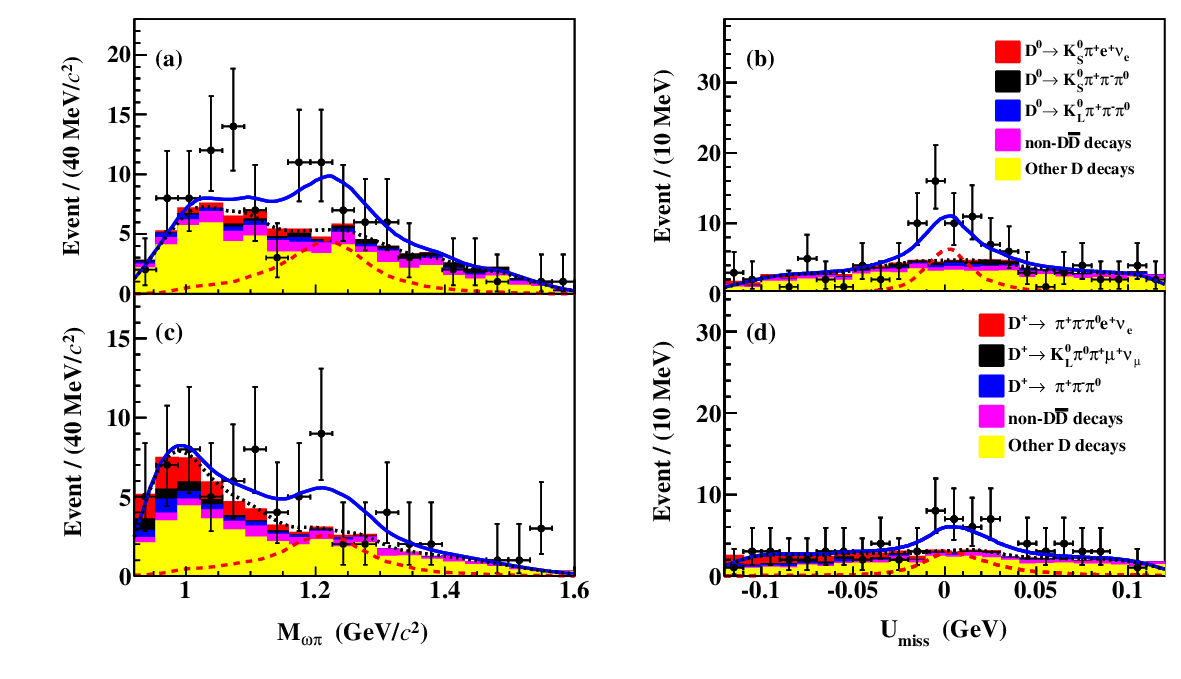}
\caption{
The projections onto (a, c)  $M_{\omega\pi^{-(0)}}$ and (b, d)  $U_{\rm miss}$ for the decays (a, b)~$D^0\to b_1^- e^+\nu_e$ and (c, d)~$D^+\to b_1^0 e^+\nu_e$. The dots with error bars are data.  The blue solid, red dashed, and black dashed
curves are the fit results, the fitted signals, and the fitted backgrounds, respectively. The color-filled histograms are from different background sources, as labeled in the legend.
}
\label{fig:fit_Umistry1}
\end{figure}

To obtain the signal yields, we perform two-dimensional (2D) unbinned maximum likelihood fits to the $M_{\omega\pi^{-(0)}}$ vs. $U_{\rm miss}$ distributions.
In the fits, the 2D signal and background shapes  are modeled by the simulated shapes derived from the signal and inclusive MC samples, respectively, and the yields of the signal and background are left free.
The projections of
the 2D fit on the $M_{_{\omega \pi^{-(0)}}}$ and $U_{\rm miss}$ distributions
are shown in  Fig.~\ref{fig:fit_Umistry1}.
From the fits, the signal yields are $N_{\rm DT}$ = $35.6\pm8.9$ and $N_{\rm DT}$ = $17.5\pm6.7$ for  $D^{0}\to b_1^- e^+\nu_e$ and  $D^{+}\to b_1^0 e^+\nu_e$, respectively, and
the  signal significances are estimated to be $5.2\sigma$ and $3.1\sigma$, by comparing the likelihoods with or without
the signal involved and taking into account the change of degrees of freedom.
Because the non-resonant component is highly suppressed in semileptonic $D$ decays and due to limited statistics, all $\omega\pi$ candidates are assumed to come from $b_1$ decays. The non-resonant component is considered as a source of systematic uncertainty.
Based on MC simulations, the detection efficiencies $\bar \varepsilon_{\rm SL}$ are estimated to be
$0.0715\pm0.0005$ and $0.0417\pm0.0004$ for $D^0\to b_1^- e^+\nu_e$ and $D^+\to b_1^0 e^+\nu_e$, respectively.

The systematic uncertainties in the branching fraction measurements are discussed below.
The uncertainty associated with the ST yield $N_{\rm ST}^{\rm tot}$ is estimated to be 0.3\%~\cite{Dtokkenu}. The uncertainty from the quoted branching fraction of $\omega\to\pi^+\pi^-\pi^0$ is 0.8\%~\cite{pdg2024}. The efficiencies of the $e^+$ tracking and PID are investigated using a control sample of $e^+e^-\to\gamma e^+e^-$~\cite{Dtokkenu},
and those of the $\pi^\pm$ tracking and PID as well as for $\pi^0$ reconstruction are estimated with control samples of $D^0\to K^-\pi^+$, $D^0\to K^-\pi^+\pi^0$, $D^0\to K^-\pi^+\pi^+\pi^-$, and $D^+\to K^-\pi^+\pi^+$ with the same tags.
The systematic uncertainties in the tracking (PID)  are assigned as 1.0\% (1.0\%) per $e^+$ and 1.0\% (1.0\%) per $\pi^\pm$, respectively.
The systematic uncertainty of the $\pi^0$ reconstruction , including photon finding, the $\pi^0$ mass window, and the 1C kinematic fit,
is assigned as 2.0\% per $\pi^0$~\cite{pi0rec}.

The systematic uncertainty associated with the $\omega$ mass window is assigned to be 1.9\% using a control sample of $D^0\to K^0_S\omega$  reconstructed versus the same $\bar D^0$ tags as  the nominal analysis. The systematic uncertainties arising from  the $E_{\rm extra\,\gamma}^{\rm max}$ and $N_{\rm extra,\pi^0}$ requirements are estimated to be 1.8\% and 2.0\% for $D^0\to b_1^- e^+\nu_e$ and $D^+\to b_1^0 e^+\nu_e$, respectively, which are estimated using the DT samples of $D^0\to K^-e^+\nu_e$ and $D^+\to K^0_Se^+\nu_e$ decays reconstructed versus the same tags as the nominal analysis.
The systematic uncertainty related to the $b_1$ resonant parameters is estimated using alternative signal MC samples, which are produced by varying the mass and width of the $b_1$ by $\pm1\sigma$. The maximum changes of the signal efficiencies, 1.4\% and 4.8\%, are assigned as the systematic uncertainties for $D^0\to b_1^- e^+\nu_e$ and $D^+\to b_1^0 e^+\nu_e$, respectively.  The uncertainty from limited MC statistics is estimated to be 0.3\%. The systematic uncertainty due to the MC generator is assigned by examining the DT efficiency with the signal MC events produced with the phase space model. The maximum changes of the DT efficiencies, 1.4\% and 0.5\%, are assigned as  the systematic uncertainties for $D^0\to b_1^-e^+\nu_{e}$ and $D^+\to b_1^0e^+\nu_{e}$, respectively.
The systematic uncertainty associated with the  cos$\theta_{\gamma,\rm miss}$ requirement
is estimated to be 1.0\% by analyzing the DT sample of $D^0\to K_{S}^{0}\pi^-e^+\nu_{e}$.
The systematic uncertainty
due to the sideband estimation of the $\omega$ background, which may be from $D^{0(+)}\to b_1^{-(0)} e^+\nu_e$ with $b_1^{-(0)}\to \pi^+\pi^-\pi^0\pi^{0(-)}$, is estimated by  examining the measured branching fractions after considering this contribution. The changes in the branching fractions,  0.5\% and 4.6\%, are taken as  the corresponding systematic uncertainties.

To estimate the systematic uncertainties of the signal and background shapes in the $U_{\rm miss}$ fit,
a simulated shape convolved with a double Gaussian function with floating parameters is chosen as the alternative signal shape,
and the simulated background shape derived from the inclusive MC sample is replaced with a 1st-order Chebyshev polynomial function.
The differences in the fitted signal yields are taken as individual systematic uncertainties. The uncertainty arising from the background shape is mainly due to unknown non-resonant decays, and is assigned
as the change of the fitted DT yield when they are fixed
by referring to the well known non-resonant fraction in
$D^+\to \bar K^*(892)^0e^+\nu_e$~\cite{kpi}.
The total systematic uncertainties are $^{+8.8\%}_{-10.7\%}$  and $^{+12.7\%}_{-13.9\%}$  by adding all the individual contributions in quadrature for $D^0\to b_1^-e^+\nu_e$ and $D^+\to b_1^0e^+\nu_e$, respectively.

The product branching fractions  are determined to be  ${\mathcal B}(D^0\to b_{1}^-e^{+}\nu_{e})\times {\mathcal B} (b^-\to \omega \pi^-) = (0.72\pm0.18^{+0.06}_{-0.08})\times10^{-4}$ and ${\mathcal B}(D^+\to b_{1}^0e^{+}\nu_{e})\times {\mathcal B} (b_1^0\to \omega \pi^0)=(1.16\pm0.44\pm0.16)\times10^{-4}$ for $D^0\to b_1^-e^+\nu_e$ and $D^+\to b_1^0e^+\nu_e$, where the first and second uncertainties are statistical and systematic, respectively.

In summary, by analyzing $7.9~\rm {fb}^{-1}$  of $e^+e^-$ collision data taken at $\sqrt{s}=3.773$~GeV,
we report the first observation of $D^0\to b_1^-e^+\nu_e$
with a  significance of 5.2$\sigma$  and find the first evidence  for $D^+\to b_1^0 e^+\nu_e$ with a  significance of 3.1$\sigma$ after considering systematic uncertainty.
Their product branching fractions  are determined to be ${\mathcal B}(D^0\to b_{1}^-e^{+}\nu_{e})\times {\mathcal B} (b_1^-\to \omega \pi^-) = (0.72\pm0.18^{+0.06}_{-0.08})\times10^{-4}$ and ${\mathcal B}(D^+\to b_{1}^0e^{+}\nu_{e})\times {\mathcal B} (b_1^0\to \omega \pi^0)=(1.16\pm0.44\pm 0.16)\times10^{-4}$.
Assuming ${\mathcal B}(b_1\to \omega\pi)=1$, these results are comparable with the theoretical predictions reported in Refs.~\cite{isgw2,cheng},
thereby implying that the $\omega\pi$ final state is currently the dominant decay mode of $b_1$, which aligns with the leading-order (LO) predictions of Refs.~\cite{Zhou:2014ila,Liang:2019vhf,Clymton:2023txd}, but conflicts with the NLO calculations from Ref.~\cite{Zhou:2014ila}.
Using the obtained branching fractions,  the world-average lifetimes of the $D^0$ and $D^+$ mesons~\cite{pdg2024}, and assuming that
${\mathcal B}(b_1^-\to \omega \pi^-)= {\mathcal B}(b_1^0 \to \omega \pi^0)$,
we determine the partial decay width ratio $\frac{\Gamma(D^0\to b_{1}^-e^{+}\nu_{e})}{2\Gamma(D^+\to b_{1}^0e^{+}\nu_{e})}=0.78\pm0.19^{+0.04}_{-0.05}$,
which is consistent with unity
within uncertainties after considering the  shared
systematic terms, and is thus consistent with
isospin conservation.
Observation of the $b_1$ in semileptonic $D$ decays opens a new avenue for experimental investigation.
Future analysis of their decay dynamics, which will be made possible with the larger data samples at BESIII~\cite{bes3-white-paper} and a potential future STCF~\cite{stcf}, will provide deep insights into the inner structure, production, mass, and width of the $b_1$,
as well as provide access to hadronic-transition form factors, and strong phase, which sheds light on the phase of elastic $\omega\pi$ scattering.

The authors appreciate Lisheng Geng, Qiang Zhao and Feng-Kun Guo for useful discussions. The BESIII Collaboration thanks the staff of BEPCII and the IHEP computing center for their strong support. This work is supported in part by National Key R\&D Program of China under Contracts Nos. 2023YFA1606000, 2023YFA1606704, 2020YFA0406300, 2020YFA0406400; National Natural Science Foundation of China (NSFC) under Contracts Nos. 12375092, 11635010, 11735014, 11935015, 11935016, 11935018, 11961141012, 12025502, 12035009, 12035013, 12061131003, 12192260, 12192261, 12192262, 12192263, 12192264, 12192265, 12221005, 12225509, 12235017; the Chinese Academy of Sciences (CAS) Large-Scale Scientific Facility Program; the CAS Center for Excellence in Particle Physics (CCEPP); Joint Large-Scale Scientific Facility Funds of the NSFC and CAS under Contract No. U1832207; 100 Talents Program of CAS; The Institute of Nuclear and Particle Physics (INPAC) and Shanghai Key Laboratory for Particle Physics and Cosmology; German Research Foundation DFG under Contracts Nos. 455635585, FOR5327, GRK 2149; Istituto Nazionale di Fisica Nucleare, Italy; Ministry of Development of Turkey under Contract No. DPT2006K-120470; National Research Foundation of Korea under Contract No. NRF-2022R1A2C1092335; National Science and Technology fund of Mongolia; National Science Research and Innovation Fund (NSRF) via the Program Management Unit for Human Resources \& Institutional Development, Research and Innovation of Thailand under Contract No. B16F640076; Polish National Science Centre under Contract No. 2019/35/O/ST2/02907; The Swedish Research Council; U. S. Department of Energy under Contract No. DE-FG02-05ER41374.

\end{document}

%% file: authorlist_2024-04-30.tex
\author{
M.~Ablikim$^{1}$\BESIIIorcid{0000-0002-3935-619X},
M.~N.~Achasov$^{4,c}$\BESIIIorcid{0000-0002-9400-8622},
P.~Adlarson$^{76}$\BESIIIorcid{0000-0001-6280-3851},
O.~Afedulidis$^{3}$\BESIIIorcid{0009-0006-2899-9946},
X.~C.~Ai$^{81}$\BESIIIorcid{0000-0003-3856-2415},
R.~Aliberti$^{35}$\BESIIIorcid{0000-0003-3500-4012},
A.~Amoroso$^{75A,75C}$\BESIIIorcid{0000-0002-3095-8610},
Q.~An$^{58,72,a}$,
Y.~Bai$^{57}$\BESIIIorcid{0000-0001-6593-5665},
O.~Bakina$^{36}$\BESIIIorcid{0009-0005-0719-7461}
% ...
\\
\vspace{0.2cm}
(BESIII Collaboration)\\
\vspace{0.2cm} {\it
$^{1}$ Institute of High Energy Physics, Beijing 100049, People's Republic of China\\
% ...
$^{3}$ Bochum  Ruhr-University, D-44780 Bochum, Germany\\
$^{4}$ Budker Institute of Nuclear Physics SB RAS (BINP), Novosibirsk 630090, Russia\\
% ...
$^{35}$ Johannes Gutenberg University of Mainz, Johann-Joachim-Becher-Weg 45, D-55099 Mainz, Germany\\
$^{36}$ Joint Institute for Nuclear Research, 141980 Dubna, Moscow region, Russia\\
% ...
$^{57}$ Southeast University, Nanjing 211100, People's Republic of China\\
$^{58}$ State Key Laboratory of Particle Detection and Electronics, Beijing 100049, Hefei 230026, People's Republic of China\\
% ...
$^{72}$ University of Science and Technology of China, Hefei 230026, People's Republic of China\\
% ...
$^{75}$ University of Turin and INFN, (A)University of Turin, I-10125, Turin, Italy; (B)University of Eastern Piedmont, I-15121, Alessandria, Italy; (C)INFN, I-10125, Turin, Italy\\
$^{76}$ Uppsala University, Box 516, SE-75120 Uppsala, Sweden\\
% ...
$^{81}$ Zhengzhou University, Zhengzhou 450001, People's Republic of China\\
% ...
$^{a}$ Deceased\\
% ...
$^{c}$ Also at the Novosibirsk State University, Novosibirsk, 630090, Russia\\
% ...
}
}